\documentclass[epj]{svjour}
\begin{onecolumn}
\usepackage{graphics}
\begin{document}
\title{Re-evaluation of neutrino mixing pattern according to latest T2K result}
\author{Ya-juan Zheng\inst{1, 2} \and Bo-Qiang Ma\inst{1,3}
\thanks{\emph{e-mail:} mabq@pku.edu.cn}%
}                     
%
%
\institute{School of Physics and State Key Laboratory of Nuclear Physics and
Technology, Peking University, Beijing 100871,
China \and Department of Physics, Shandong University, Jinan, Shandong 250100, China \and Center for High Energy
Physics, Peking University, Beijing 100871,
China}
\date{Received: date / Revised version: date}
%
\abstract{
We re-evaluate neutrino mixing patterns according to the latest T2K
result for a larger mixing angle $\theta_{13}$, and find that the
PMNS mixing matrix has larger deviations from bimaximal (BM) and
tribimaximal (TB) mixing patterns than previously expected. We also
find that several schemes connecting PMNS and CKM mixing matrices
can accommodate the latest T2K result nicely. As necessary updates
to former works, we make new triminimal expansions of PMNS mixing
matrix based on BM and TB mixing patterns. We also propose a new
mixing pattern with a self-complementary relation between the mixing
angles $\theta_{12}^{\nu} + \theta_{13}^{\nu} \simeq 45^\circ$,
and find such a new mixing pattern in leading order can provide a
rather good description of the data.
} 
\maketitle

\section{Introduction}
\label{intro}
Neutrino oscillations observed in the solar~\cite{SNO},
atmospheric~\cite{SK}, reactor~\cite{KM} and accelerator~\cite{K2K}
neutrino experiments have provided plentiful information on neutrino
masses and neutrino mixings. Among the knowns and unknowns of
neutrino physics, the nonzero but nontrivial smallest mixing angle
$\theta_{13}$ has received a lot of attentions, therefore the
measurement of $\theta_{13}$ is important to enrich our
understanding of neutrino properties. It may lead to the measurement
of CP violation in the lepton sector and the study of its origin. In
addition, the precision measurement of $\theta_{13}$ would greatly
enhance our knowledge to the solar, atmospheric, the secondary
effects of long baseline, and the supernova neutrino oscillations as
well.

In the last decade, international collaborations have been
established to focus on the measurement of $\theta_{13}$, e.g.,
Double Chooz in France, Reactor Experiment for Neutrino Oscillations
(RENO) in South Korea, Daya Bay Experiment in China, NO$\nu$A in
United States, and Tokai to Kamioka (T2K) experiment in Japan. With
the appearance of the latest result from T2K~\cite{t2k}, which
indicates a relatively large $\theta_{13}$, we need to re-evaluate
our understanding on neutrino masses and mixings, especially on the
neutrino mixing pattern~\cite{xing&qinma1,xing&qinma2}.

Ever since the establishment of neutrino oscillation, theorists have
been working on the explanation of neutrino flavor mixings. From
bimaximal (BM) mixing, trimaximal (TB) mixing to the tribimaximal
mixing, various neutrino mixing patterns have been brought up and
later evaluated by the experimental data.  On the other hand, the
underlying unification and symmetry concerning leptons and quarks
are speculated from some clues of low energy experimental data, such
as the quark-lepton complementarity (QLC)~\cite{QLC1,QLC2,QLC3,QLC4,QLC5,QLC6,QLC7,QLC8,QLC9,QLC10,QLC11,QLC12}. As the first
result from T2K offers new bounds on the smallest neutrino mixing
angles, we aim to make an analysis on the implications and
consequences from the observation of a nonzero and also not so small
value of $\theta_{13}$.

The remaining part of this paper is organized as follows. In
Sec.~\ref{section:exp}, we check experimental status of quark and
lepton mixings based on previous fits together with the latest T2K
result of a larger mixing angle $\theta_{13}$. It is remarkable that
the deviations of the lepton mixing matrix from bimaximal and
tribimaximal mixing patterns are expressed in terms of the
Wolfenstein parameter $\lambda$ of quark mixing matrix. Then we
examine in Sec.~\ref{section:relating} several schemes connecting
the mixing matrices of quarks and leptons, and find that some of
them are possible to accommodate the latest T2K data. In
Sec.~\ref{section:triminimal}, as a necessary update, we make new
expansions of PMNS mixing matrix based on both bimaximal and
tribimaximal mixing patterns, with a more reasonable hierarchy
structure in powers of the Wolfenstein parameter $\lambda$ in the
quark mixing matrix. In Sec.~\ref{section:new} from the
phenomenology point of view, we propose a new mixing pattern with a
self-complementary relation between the mixing angles
$\theta_{12}^{\nu} + \theta_{13}^{\nu} \simeq 45^\circ$. The new
mixing matrix is more closer to the data than previously known
matrices such as BM and TB mixing patterns. Then we provide a
summary in Sec.~\ref{section:summary}.

\section{Experimental status of quark and lepton mixings}\label{section:exp}

Quark and lepton mixings are phenomenologically described by the
Cabibbo-Kobayashi-Maskawa~(CKM)~\cite{CKM1,CKM2} matrix and
Pontecorvo-Maki-Nakagawa-Sakata~(PMNS)~\cite{PMNS1,PMNS2} matrix
respectively. The standard parameterization~\cite{ck} of quark and
lepton mixing matrices is expressed by three mixing angles
$\theta_{12}$, $\theta_{13}$, $\theta_{23}$ and one CP-violating
phase angle $\delta$. As shown below, the elements of the mixing
matrix in the first row and third column adopt very simple forms
\begin{eqnarray}
V = \left (
\begin{array}{ccc}
c^{~}_{12} c_{13}         & s^{~}_{12} c_{13}       & s_{13} e^{-i\delta}\cr
-c^{~}_{12} s_{23} s_{13} e^{i\delta}- s^{~}_{12} c_{23}
& -s^{~}_{12} s_{23} s_{13}e^{i\delta} + c^{~}_{12} c_{23}   & s_{23} c_{13} \cr
-c^{~}_{12} c_{23} s_{13} e^{i\delta}+ s^{~}_{12} s_{23}
& -s^{~}_{12} c_{23} s_{13} e^{i\delta}- c^{~}_{12} s_{23}   & c_{23} c_{13}
\end{array}
\right),
\end{eqnarray}
 where $s_{ij}={\rm sin}\theta_{ij}, c_{ij}={\rm cos}\theta_{ij}~(i,j=1,2,3)$.
 For describing the quark mixing matrix $V_{\rm CKM}$, another simple form of parametrization, i.e., the Wolfenstein parametrization was proposed with the defination that $s_{12}=\lambda,~s_{23}=A\lambda^2$ and $s_{13}e^{i\delta}=A\lambda^3(\rho+i\eta)$~\cite{Wolfenstein} . Its explicit form at the accuracy of ${\cal O}(\lambda^4)$ is
\begin{equation}
V_{\rm CKM} = \left (
\begin{array}{ccc}
1-\frac{1}{2}\lambda^{2}    & \lambda   & A\lambda^{3}(\rho
-i\eta) \\
-\lambda    &
1-\frac{1}{2}\lambda^{2}    &
A\lambda^{2} \\
A\lambda^{3}(1-\rho-i\eta)  & -A\lambda^{2}     & 1
\end{array}
\right )+{\cal O}(\lambda^4) \; ,
\end{equation}
and we adopt the following inputs given by the
Particle Data Group \cite{PDG}:
\begin{eqnarray}
\lambda = 0.2257^{+0.0009}_{-0.0010},~~~~A=0.814^{+0.021}_{-0.022},~~~~\bar{\rho}=0.135^{+0.031}_{-0.016},~~~~~~\bar{\eta}=0.349^{+0.015}_{-0.017},
\end{eqnarray}
where $\bar{\rho}=\rho - \frac{1}{2}\rho \lambda^2 +O(\lambda^4)$
and $\bar{\eta} =\eta-\frac{1}{2}\eta \lambda^2 +O(\lambda^4)$. The
corresponding quark mixing angles could be obtained by calculating
the moduli of the mixing matrix elements
$\theta_{12}^{q}=13.04^{\circ
+0.053^\circ}_{~-0.059^\circ},\theta_{23}^q=2.37^{\circ+0.081^\circ}_{~-0.085^\circ}$
and
$\theta_{13}^q=0.20^{\circ+0.023^\circ}_{~-0.020^\circ}$~\cite{zheng}.
One advantage of the Wolfenstein parametrization is that one may
estimate the order of magnitude of any element from the hierarchical
feature of $\lambda$. It is natural to recall that the deviation of
$V_{\rm CKM}$ can be considered as expansions in orders of $\lambda$
from the unit matrix $I$. To assume firstly that the $V_{\rm CKM}$
quark mixing matrix takes the form of unit matrix, it is easy to get
three $0^\circ$ quark mixing angles in the unit mixing pattern.
Hence, we can obtain the corresponding deviation of each mixing
angle:
\begin{eqnarray}
\left\{\begin{array}{l}
\Delta_{23}^{I}=|\theta_{23}^{q}-\theta_{23}^{I}|\simeq 2.37^{\circ},~~~~~{\rm sin}\Delta_{23}^{I}\sim~{\cal O}(\lambda^2),\\
\Delta_{12}^{I}=|\theta_{12}^{q}-\theta_{12}^{I}|\simeq 13.04^{\circ},~~~{\rm sin}\Delta_{12}^{I}\sim~{\cal O}(\lambda),\\
\Delta_{13}^{I}=|\theta_{13}^{q}-\theta_{13}^{I}|\simeq 0.20^{\circ},~~~~~{\rm sin}\Delta_{13}^{I}\sim~{\cal O}(\lambda^3),
\end{array}\right.
\end{eqnarray}
where $\lambda\simeq 0.2$ denotes the Wolfenstein parameter.

For the neutrino sector, with the latest global fit of experimental data given in~\cite{nu_data1,nu_data2}, the three mixing angles of the PMNS mixing matrix $U_{\mathrm{PMNS}}$ read as
\begin{eqnarray}
\nonumber
{\rm sin}^2\theta_{12}=0.312(1^{+0.128}_{-0.109})~(2\sigma),\\
\nonumber
{\rm sin}^2\theta_{23}=0.466(1^{+0.292}_{-0.215}) ~(2\sigma), \\
{\rm sin}^2\theta_{13}=0.016\pm 0.010 ~(1\sigma),
\end{eqnarray}
which indicate that
\begin{eqnarray}
\theta_{12}^{\nu}\simeq 33.957^{\circ +2.434^\circ}_{~-2.143^\circ},~~~\theta_{23}^{\nu}\simeq 43.050^{\circ +7.839^\circ}_{~-5.834^\circ},~~~{\rm and}~~~\theta_{13}^{\nu}\simeq 7.27^{\circ +2.012^\circ}_{~-2.824^\circ},
\end{eqnarray}
but the CP violating phase $\delta$ remains unconstrained. As the
recent T2K collaboration results~\cite{t2k} give more robust
indication of a relatively large $\theta_{13}$ for $\delta_{\rm
CP}=0$:
\begin{eqnarray}
\nonumber
0.03< {\rm sin}^22\theta_{13}<0.28,~~~~~~~~~~~{\rm Normal~~ Hierarchy~(NH)},\\
0.04<{\rm sin}^22\theta_{13}<0.34,~~~~~~~~~~{\rm Inverted ~~Hierarchy~(IH)},
\end{eqnarray}
the smallest mixing angle $\theta_{13}$ could be naturally produced
\begin{eqnarray}
4.987^\circ <\theta_{13}^{\nu}<15.974^\circ~~({\rm NH}),~~~~ 5.769^\circ<\theta_{13}^{\nu}<17.834^\circ~~({\rm IH}).
\end{eqnarray}
The best-fit values are ${\rm sin}^22\theta_{13}=0.11~({\rm
NH})~{\rm or}~{\rm sin}^22\theta_{13}=0.14~({\rm IH})$, i.e.,
\begin{eqnarray}
\theta_{13}^{\nu}=9.685^{\circ +4.698^\circ}_{~-6.289^\circ}~~{(\rm NH})~~~~{\rm or}~~\theta_{13}^{\nu}=10.986^{\circ+5.218^\circ}_{~-6.848^\circ}~~({\rm IH}).
\end{eqnarray}

In recent years, lots of efforts have been devoted to explore a new
theoretical framework to accommodate tiny neutrino masses and large
flavor mixing angles. Among those studies which have tried to
parametrize the PMNS matrix with only constant numbers, the proposal
of bimaximal (BM)~\cite{bi1,bi2,bi3,bi4,bi5,bi6} and tribimaximal (TB)~\cite{tri1,tri2,tri3,tri4,tri5} mixing
patterns are considered to be the most successful parametrizations
and are mostly discussed. The specific forms are written as
\begin{eqnarray}
U_{\rm BM}=\left(
\begin{array}{ccc}
\sqrt{2}/2& \sqrt{2}/2 & 0 \\
-1/2 & 1/2 & \sqrt{2}/2 \\
1/2 & -1/2 & \sqrt{2}/2
\end{array}\right)P_\nu,
\end{eqnarray}
\begin{eqnarray}
U_{\rm TB}=\left(
\begin{array}{ccc}
2/\sqrt{6}& 1/\sqrt{3} & 0 \\
-1/\sqrt{6} & 1/\sqrt{3} & 1/\sqrt{2} \\
1/\sqrt{6} &  -1/\sqrt{3}& 1/\sqrt{2}
\end{array}\right)P_\nu,
\label{TBform}
\end{eqnarray}
where $P_\nu={\rm Diag}\{e^{-i\alpha/2},e^{-i\beta/2},1\}$ includes
two CP-violating phases if the three neutrinos are Majorana
fermions. These two scenarios give the prediction of neutrino mixing
angles
\begin{eqnarray}
\nonumber
\theta_{23}^{\rm BM}=45^\circ,~~~\theta_{12}^{\rm BM}=45^\circ,~~~ \theta_{13}^{\rm BM}=0^\circ,\\
\theta_{23}^{\rm TB}=45^\circ,~~~\theta_{12}^{\rm TB}=35.26^\circ, ~~~\theta_{13}^{\rm TB}=0^\circ,
\end{eqnarray}
respectively in the standard parametrization of the 3$\times$3
neutrino mixing matrix. In these two cases the Dirac CP-violating
phase $\delta$ is absent, which leads to no CP violation in neutrino
oscillations.

With the above experimental results as well as phenomenological
deduction, combined with the quark-lepton complementarity (QLC)
relations~\cite{QLC1,QLC2,QLC3,QLC4,QLC5,QLC6,QLC7,QLC8,QLC9,QLC10,QLC11,QLC12} given by
\begin{eqnarray}
\theta_{12}^q+\theta_{12}^{\nu}=45^\circ,\quad\theta_{23}^q+\theta_{23}^{\nu}=45^\circ,
\label{qlc}
\end{eqnarray}
we can directly compare the deviation of PMNS lepton mixing matrix
from BM and TB mixing patterns in terms of three mixing angles:
\begin{eqnarray}
\left\{\begin{array}{l}
\Delta_{23}^{\rm BM}=|\theta_{23}^{\nu}-\theta_{23}^{\rm BM}|\simeq 2.950^{\circ},~~~~~~~~~~~~~~~{\rm sin}\Delta_{23}^{\rm BM}\sim~{\cal O}(\lambda^2),\\
\Delta_{12}^{\rm BM}=|\theta_{12}^{\nu}-\theta_{12}^{\rm BM}|\simeq 11.043^{\circ},~~~~~~~~~~~~~~{\rm sin}\Delta_{12}^{\rm BM}\sim~{\cal O}(\lambda),\\
\Delta_{13}^{\rm BM}=|\theta_{13}^{\nu}-\theta_{13}^{\rm BM}|\simeq 9.685^\circ~(10.986^\circ) ,~~~{\rm sin}\Delta_{13}^{\rm BM}\sim~{\cal O}(\lambda),
\end{array}\right.
\end{eqnarray}
and
\begin{eqnarray}
\left\{\begin{array}{l}
\Delta_{23}^{\rm TB}=|\theta_{23}^{\nu}-\theta_{23}^{\rm TB}|\simeq 2.950^{\circ},~~~~~~~~~~~~~~~~{\rm sin}\Delta_{23}^{\rm TB}\sim~{\cal O}(\lambda^2),\\
\Delta_{12}^{\rm TB}=|\theta_{12}^{\nu}-\theta_{12}^{\rm TB}|\simeq 1.303^{\circ},~~~~~~~~~~~~~~~~{\rm sin}\Delta_{12}^{\rm TB}\sim~{\cal O}(\lambda^3),\\
\Delta_{13}^{\rm TB}=|\theta_{13}^{\nu}-\theta_{13}^{\rm TB}|\simeq 9.685^\circ~(10.986^\circ),~~~{\rm sin}\Delta_{13}^{\rm TB}\sim~{\cal O}(\lambda).
\end{array}\right.
\end{eqnarray}
From above we find that the mixing angle $\Delta_{13}^{\nu} \sim
{\cal O}(\lambda)$, which serves as a larger deviation from BM and
TB mixing patterns according to the latest T2K result of larger
$\theta_{13}$ for neutrino oscillation.  This situation  is
different from our expectation $\Delta_{13}^{\nu} \sim {\cal
O}(\lambda^2)$ or $(\lambda^3)$ with small $\theta_{13}$, which make
it necessary to update the corresponding
study~\cite{triminimalquark2} on expansions based on the deviation.

\subsection{Relating quarks and leptons}
\label{section:relating}

In this section, we aim to explore the seemingly independent mixing
patterns between quark and lepton sectors. The PMNS matrix depends
generally on the charged lepton sector whose diagonalization leads
to a charged lepton mixing matrix $U_l^\dagger$ which should be
combined with the neutrino mixing matrix $U_\nu$~\cite{Minakata:2004xt1,Minakata:2004xt2}:
\begin{eqnarray}
U_{\rm PMNS}=U_l^\dagger U_\nu .
\end{eqnarray}

Since the CKM matrix is quite near unit matrix, and the PMNS
matrix has been considered to be close to bimaximal matrix and
tribimaximal mixing matrix, it is interesting to assume that the
deviation of the PMNS matrix from the BM and TB mixings can be
described by the CKM matrix as generally discussed in reference~\cite{Minakata:2004xt1,Minakata:2004xt2}, that is,
\begin{eqnarray}
U_{\rm PMNS}V_{\rm CKM}=U_{\rm BM/TB},~~~{\rm or}~~~ V_{\rm
CKM}U_{\rm PMNS}=U_{\rm BM/TB}, \label{UVrelation}
\end{eqnarray}
which is equivalent to
\begin{eqnarray}
U_{\rm PMNS}=U_{\rm BM/TB}V_{\rm CKM}^\dagger,~~~{\rm or} ~~~U_{\rm
PMNS}=V_{\rm CKM}^\dagger U_{\rm BM/TB} .
\end{eqnarray}

Adopting similar interpretation, Ref.~\cite{linan} provides explicit
discussions of the following two cases for the relation between CKM
and PMNS mixing matrices with $U_\nu$ the bimaximal mixing matrix:
\begin{eqnarray}
U_{\rm PMNS}=U_{\rm BM}V_{\rm CKM}^\dagger, ~~{\rm and} ~~U_{\rm
PMNS}=V_{\rm CKM}^\dagger U_{\rm BM}.
\end{eqnarray}
In the first case, substituting the corresponding matrices of $U_{\rm BM}$ and $V_{\rm CKM}$ one can calculate the moduli of each mixing element of $U_{\rm PMNS}$ and give
\begin{eqnarray}
|s_{13}^{\rm PMNS}|=\frac{\sqrt{2}}{2}A\lambda^2\sqrt{(\lambda-\lambda \rho-1)^2+(\lambda \eta)^2}=0.48\lambda^2.
\end{eqnarray}
Thus, we get the explicit value
\begin{eqnarray}
1.39^\circ<\theta_{13}<1.40^\circ,
\end{eqnarray}
assuming no CP violation. It is obvious that this region is excluded by the T2K results. Thus this scheme is not preferable judging by present experimental data.

In the second case, $U_{13}$ could be produced similarly
\begin{eqnarray}
|s_{13}^{\rm PMNS}|=\frac{\sqrt{2}}{2}\lambda\sqrt{[A\lambda^2(1-\rho)-1]^2+(A\lambda^2\eta)^2}=0.68\lambda,
\end{eqnarray}
and consequently
\begin{eqnarray}
8.79^\circ<\theta_{13}<8.84^\circ.
\end{eqnarray}
Although the second case provides $\theta_{13}$ values more closer
to the experimental data, it is proved in the original paper that this case does not
accommodate the QLC relations, i.e., Eq.~(\ref{qlc}), very
well~\cite{linan}.

We should point out that the multiplying relationship between PMNS
and CKM matrices, such as Eq.~(\ref{UVrelation}), is not independent
from different parametrizations of the charged lepton CKM matrix.
The reason is that different choices of phase factors in $V_{\rm
CKM}$ can produce differences in the predicted $U_{\rm PMNS}$.
 If we choose the leptonic mixing matrix
$U_\nu$ to be tribimaximal pattern, for another choice of  charged
lepton mixing matrix $U^{\dagger}_l$ in the case of $U_{\rm
PMNS}=U_{l}^\dagger U_{\rm TB}$, an {\it Ansatz} have been
discussed~\cite{cheng1,cheng2} using a new
parametrization~\cite{qm_parametrization}:
 \begin{eqnarray}
  U^{\dagger}_{l}={\left(\begin{array}{ccc}
   1-\lambda^2/2 & \lambda e^{i\delta} & h\lambda^3 \\
   -\lambda e^{-i\delta} & 1-\lambda^2/2 & (f+h e^{-i\delta})\lambda^2 \\
   f\lambda^3 e^{-i\delta} & -(f+h e^{i\delta})\lambda^2 & 1 \\
   \end{array}\right)}+{\cal O}(\lambda^{4})~,
 \label{Vl}
 \end{eqnarray}
where the parameters $A$, $\rho$ and $\eta$ in the Wolfenstein
parametrization~\cite{Wolfenstein} are replaced by $f$, $h$ and
$\delta$ in the new Wolfenstein-like one~\cite{qm_parametrization}.
And the corresponding parameters read as:
 \begin{eqnarray}
 f=0.749^{+0.034}_{-0.037}\,,\quad h=0.309^{+0.017}_{-0.012}\,,\quad \lambda=0.22545\pm 0.00065\,,
 \quad \delta=(89.6^{+2.94}_{-0.86})^\circ\,.
   \end{eqnarray}
 As we will see below, this matrix could be a candidate for the small deviation of the PMNS matrix from the TB mixing pattern.

 By substituting the corresponding matrix in $U_{\rm
PMNS}=U^\dagger_l U_\nu$, the leptonic mixing matrix corrected by
the contributions from $U_l^\dagger$ can be written, up to order of
$\lambda^{3}$, as~\cite{cheng1,cheng2}
 \begin{eqnarray}
  U_{\rm PMNS}&=&U_{\rm TB}+{\left(\begin{array}{ccc}
   -\frac{\lambda e^{i\delta}}{\sqrt{6}}-\frac{\lambda^{2}(1+h\lambda)}{\sqrt{6}} & \frac{\lambda e^{i\delta}}{\sqrt{3}}-\frac{\lambda^{2}(1-2h\lambda)}{2\sqrt{3}} & \frac{\lambda(h\lambda^{2}-e^{i\delta})}{\sqrt{2}} \\
   -\lambda\sqrt{\frac{2}{3}} e^{-i\delta}-\frac{\lambda^2(1-2f-2he^{-i\delta})}{2\sqrt{6}} & -\frac{\lambda e^{-i\delta}}{\sqrt{3}}-\frac{\lambda^{2}(1-2f-2h\lambda e^{-i\delta})}{2\sqrt{3}} & \frac{\lambda^2(1+2f+2h e^{-i\delta})}{2\sqrt{2}} \\
   \frac{\lambda^2(f+h e^{i\delta}+2f\lambda e^{-i\delta})}{\sqrt{6}}  & -\frac{\lambda^2(f+h e^{i\delta}-f\lambda e^{-i\delta})}{\sqrt{3}} & \frac{\lambda^2(f+h e^{i\delta})}{\sqrt{2}} \\
   \end{array}\right)}P_{\nu} \nonumber\\
   &+&{\cal O}(\lambda^4) \ .
 \end{eqnarray}
In this case, Ref.~\cite{cheng1} adopts an alternative explicit form
for tribimaximal mixing matrix with negative 21, 23 and 31 matrix
elements, which is physically equivalent to Eq.~(\ref{TBform}).
Thus, the smallest mixing angle $\theta_{13}$ can be obtained
by~\cite{cheng1}
\begin{eqnarray}
  \sin\theta_{13}&=&\frac{\lambda}{\sqrt{2}}(1-h\lambda^{2}\cos\delta)~,\nonumber
\end{eqnarray}
with a non-vanishing $\theta_{13}=9.2^{\circ}$. Such a scheme, which
has been proposed by Ahn-Cheng-Oh~\cite{cheng1,cheng2}, provides a viable
relation to connect quark and lepton mixing matrices in agreement
with the new T2K results.

\section{ Triminimal expansion}\label{section:triminimal}

After the above initiative study of T2K results, we found that the
relatively large $\theta_{13}$ makes it necessary to update and
expand the former works on triminimal
expansion~\cite{triminimalquark,triminimalquark2}. If the flavor
mixing matrices are parametrized according to the hierarchical
structure of mixing, it may reveal more physical information about
the underlying theory. A good choice of this idea is the triminimal
parametrization~\cite{triminimalneutrino1,triminimalneutrino2,triminimalquark,triminimalquark2}
with an approximation as the basis matrix to the lowest order. That
is to express a mixing angle in the mixing matrix as the sum of a
zeroth order angle $\theta^0$ and a small perturbation angle
$\epsilon$ with
\begin{eqnarray}
\theta_{12}=\theta_{12}^0+\epsilon_{12},\quad\theta_{23}=\theta_{23}^0+\epsilon_{23},\quad\theta_{13}=\theta_{13}^0+\epsilon_{13}.
\end{eqnarray}
With the deviations $\epsilon_{ij}$, one can expand the mixing matrix in powers of $\epsilon_{ij}$ while different choices
of $\theta_{ij}^0$ lead to different basis.

 According to new results of neutrino mixing angles, we can see that
the deviation of PMNS mixing matrix from TB mixing pattern is more
close to that of CKM mixing matrix from unit matrix if measured in
terms of mixing angles. To be explicit, the comparison of deviation
hierarchy is listed as follows for corresponding mixing angles:
${\rm sin}\Delta_{13}^{\rm TB}\sim{\rm sin}\Delta_{12}^{I}\sim{\cal
O}(\lambda)$ and ${\rm sin}\Delta_{23}^{\rm TB}\sim{\rm
sin}\Delta_{23}^{I}\sim{\cal O}(\lambda^2)$.  From this aspect, the
deviation hierarchy from BM and TB mixing patterns of different
mixing angles can serve as the triminimal expansion basis in
constructing the PMNS matrix.

\subsection{Expansion on bimaximal pattern}

To realize the triminimal expansion on BM mixing pattern, we assume
that $\epsilon^L_{12}=-\lambda$ and $\epsilon^L_{23}=-A\lambda^2$
considering the suggestion of QLC~\cite{triminimalquark2}, and
$\epsilon^L_{13}e^{i\delta^L}=\lambda Z$ from the constraint of new
T2K result as estimated in Sec.~\ref{section:exp}. To the third order in
$\epsilon^L_{ij}$, the expansion is~\cite{triminimalquark2}

\begin{eqnarray}
U_{\rm PMNS}&=&U_{\rm BM}
+ \epsilon^L_{12}\left(\begin{array}{ccc}  -\frac{\sqrt{2}}{2} & \frac{\sqrt{2}}{2} & 0 \\
                                           -\frac{1}{2} & -\frac{1}{2} & 0 \\
                                           \frac{1}{2} & \frac{1}{2} & 0 \\
                                           \end{array}
                                           \right)
  +\epsilon^L_{23}\left(
                  \begin{array}{ccc}
                    0 & 0 & 0 \\
                    \frac{1}{2} & -\frac{1}{2} & \frac{\sqrt{2}}{2} \\
                    \frac{1}{2} & -\frac{1}{2} & -\frac{\sqrt{2}}{2} \\
                  \end{array}
                \right)
  +\epsilon^L_{13}\left(
                  \begin{array}{ccc}
                    0 & 0 & e^{-i\delta^L} \\
                    -\frac{1}{2}e^{i\delta^L} & -\frac{1}{2}e^{i\delta^L} & 0 \\
                    -\frac{1}{2}e^{i\delta^L} & -\frac{1}{2}e^{i\delta^L} & 0 \\
                  \end{array}
                \right)\nonumber\\
  &+&(\epsilon^L_{12})^2\left(
                  \begin{array}{ccc}
                    -\frac{\sqrt{2}}{4} & -\frac{\sqrt{2}}{4} & 0 \\
                    \frac{1}{4} & -\frac{1}{4} & 0 \\
                    -\frac{1}{4} & \frac{1}{4} & 0 \\
                  \end{array}
                \right)
  +(\epsilon^L_{23})^2\left(
                  \begin{array}{ccc}
                    0 & 0 & 0 \\
                    \frac{1}{4} & -\frac{1}{4} & -\frac{\sqrt{2}}{4} \\
                    -\frac{1}{4} & \frac{1}{4} & -\frac{\sqrt{2}}{4} \\
                  \end{array}
                \right)
  +(\epsilon^L_{13})^2\left(
                  \begin{array}{ccc}
                    -\frac{\sqrt{2}}{4} & -\frac{\sqrt{2}}{4} & 0 \\
                    0 & 0 & -\frac{\sqrt{2}}{4} \\
                    0 & 0 & -\frac{\sqrt{2}}{4} \\
                  \end{array}
                \right)\nonumber\\
  &+&\epsilon^L_{12}\epsilon^L_{23}\left(
                  \begin{array}{ccc}
                    0 & 0 & 0 \\
                    \frac{1}{2} & \frac{1}{2} & 0 \\
                    \frac{1}{2} & \frac{1}{2} & 0 \\
                  \end{array}
                \right)
  +\epsilon^L_{12}\epsilon^L_{13}e^{i\delta^L}\left(
                  \begin{array}{ccc}
                    0 & 0 & 0 \\
                    \frac{1}{2} & -\frac{1}{2} & 0 \\
                    \frac{1}{2} & -\frac{1}{2} & 0 \\
                  \end{array}
                \right)
  +\epsilon^L_{23}\epsilon^L_{13}e^{i\delta^L}\left(
                  \begin{array}{ccc}
                    0 & 0 & 0 \\
                    -\frac{1}{2} & -\frac{1}{2} & 0 \\
                    \frac{1}{2} & \frac{1}{2} & 0 \\
                  \end{array}
                \right)\nonumber\\
  &+&(\epsilon^L_{12})^3\left(
                  \begin{array}{ccc}
                    \frac{\sqrt{2}}{12} & -\frac{\sqrt{2}}{12} & 0 \\
                    \frac{1}{12} & \frac{1}{12} & 0 \\
                    -\frac{1}{12} & -\frac{1}{12} & 0 \\
                  \end{array}
                \right)
  +(\epsilon^L_{23})^3\left(
                  \begin{array}{ccc}
                    0 & 0 & 0 \\
                    -\frac{1}{12} & \frac{1}{12} & -\frac{\sqrt{2}}{12} \\
                    -\frac{1}{12} & \frac{1}{12} & \frac{\sqrt{2}}{12} \\
                  \end{array}
                \right)
  +(\epsilon^L_{13})^3\left(
                  \begin{array}{ccc}
                    0 & 0 & -\frac{1}{6}e^{-i\delta^L} \\
                    \frac{1}{12}e^{i\delta^L} & \frac{1}{12}e^{i\delta^L} & 0 \\
                    \frac{1}{12}e^{i\delta^L} & \frac{1}{12}e^{i\delta^L} & 0 \\
                  \end{array}
                \right)\nonumber\\
  &+&\epsilon^L_{12}(\epsilon^L_{23})^2\left(
                  \begin{array}{ccc}
                    0 & 0 & 0 \\
                    \frac{1}{4} & \frac{1}{4} & 0 \\
                    -\frac{1}{4} & -\frac{1}{4} & 0 \\
                  \end{array}
                \right)
  + (\epsilon^L_{12})^2\epsilon^L_{23}\left(
                  \begin{array}{ccc}
                    0 & 0 & 0 \\
                    -\frac{1}{4} & \frac{1}{4} & 0 \\
                    -\frac{1}{4} & \frac{1}{4} & 0 \\
                  \end{array}
                \right)
  +\epsilon^L_{12}(\epsilon^L_{13})^2\left(
                  \begin{array}{ccc}
                    \frac{\sqrt{2}}{4} & -\frac{\sqrt{2}}{4} & 0 \\
                    0 & 0 & 0 \\
                    0 & 0 & 0 \\
                  \end{array}
                \right)\nonumber\\
  &+&(\epsilon^L_{12})^2\epsilon^L_{13}e^{i\delta^L}\left(
                  \begin{array}{ccc}
                    0 & 0 & 0 \\
                    \frac{1}{4} & \frac{1}{4} & 0 \\
                    \frac{1}{4} & \frac{1}{4} & 0 \\
                  \end{array}
                \right)
  +\epsilon^L_{23}(\epsilon^L_{13})^2\left(
                  \begin{array}{ccc}
                    0 & 0 & 0 \\
                    0 & 0 & -\frac{\sqrt{2}}{4} \\
                    0 & 0 & \frac{\sqrt{2}}{4} \\
                  \end{array}
                \right)
  +(\epsilon^L_{23})^2\epsilon^L_{13}e^{i\delta^L}\left(
                  \begin{array}{ccc}
                    0 & 0 & 0 \\
                    \frac{1}{4} & \frac{1}{4} & 0 \\
                    \frac{1}{4} & \frac{1}{4} & 0 \\
                  \end{array}
                \right)\nonumber\\
  &+&\epsilon^L_{12}\epsilon^L_{23}\epsilon^L_{13}e^{i\delta^L}\left(
                  \begin{array}{ccc}
                    0 & 0 & 0 \\
                    \frac{1}{2} & -\frac{1}{2} & 0 \\
                    -\frac{1}{2} & \frac{1}{2} & 0 \\
                  \end{array}
                \right)+\mathcal{O}\left((\epsilon^L_{ij})^4\right).\label{exppan}
\end{eqnarray}
Since the hierarchy is $\epsilon_{23}^L\sim(\epsilon_{12}^L)^2$,
$\epsilon_{13}e^{i\delta^L}\sim \epsilon_{12}^L$, then to the third
order in $\epsilon_{12}^L$, we have
\begin{eqnarray}
U_{\rm PMNS}&=&U_{\rm BM}
+ \epsilon^L_{12}\left(\begin{array}{ccc}  -\frac{\sqrt{2}}{2} & \frac{\sqrt{2}}{2} & 0 \\
                                           -\frac{1}{2} & -\frac{1}{2} & 0 \\
                                           \frac{1}{2} & \frac{1}{2} & 0 \\
                                           \end{array}
                                           \right)
  +\epsilon^L_{23}\left(
                  \begin{array}{ccc}
                    0 & 0 & 0 \\
                    \frac{1}{2} & -\frac{1}{2} & \frac{\sqrt{2}}{2} \\
                    \frac{1}{2} & -\frac{1}{2} & -\frac{\sqrt{2}}{2} \\
                  \end{array}
                \right)
  +\epsilon^L_{13}\left(
                  \begin{array}{ccc}
                    0 & 0 & e^{-i\delta^L} \\
                    -\frac{1}{2}e^{i\delta^L} & -\frac{1}{2}e^{i\delta^L} & 0 \\
                    -\frac{1}{2}e^{i\delta^L} & -\frac{1}{2}e^{i\delta^L} & 0 \\
                  \end{array}
                \right)\nonumber\\
  &+&(\epsilon^L_{12})^2\left(
                  \begin{array}{ccc}
                    -\frac{\sqrt{2}}{4} & -\frac{\sqrt{2}}{4} & 0 \\
                    \frac{1}{4} & -\frac{1}{4} & 0 \\
                    -\frac{1}{4} & \frac{1}{4} & 0 \\
                  \end{array}
                \right)
  +(\epsilon^L_{13})^2\left(
                  \begin{array}{ccc}
                    -\frac{\sqrt{2}}{4} & -\frac{\sqrt{2}}{4} & 0 \\
                    0 & 0 & -\frac{\sqrt{2}}{4} \\
                    0 & 0 & -\frac{\sqrt{2}}{4} \\
                  \end{array}
                \right)
  +\epsilon^L_{12}\epsilon^L_{23}\left(
                  \begin{array}{ccc}
                    0 & 0 & 0 \\
                    \frac{1}{2} & \frac{1}{2} & 0 \\
                    \frac{1}{2} & \frac{1}{2} & 0 \\
                  \end{array}
                \right)
 +\epsilon^L_{12}\epsilon^L_{13}e^{i\delta^L}\left(
                  \begin{array}{ccc}
                    0 & 0 & 0 \\
                    \frac{1}{2} & -\frac{1}{2} & 0 \\
                    \frac{1}{2} & -\frac{1}{2} & 0 \\
                  \end{array}
                \right)\nonumber\\
 & +&\epsilon^L_{23}\epsilon^L_{13}e^{i\delta^L}\left(
                  \begin{array}{ccc}
                    0 & 0 & 0 \\
                    -\frac{1}{2} & -\frac{1}{2} & 0 \\
                    \frac{1}{2} & \frac{1}{2} & 0 \\
                  \end{array}
                \right)
  +(\epsilon^L_{12})^3\left(
                  \begin{array}{ccc}
                    \frac{\sqrt{2}}{12} & -\frac{\sqrt{2}}{12} & 0 \\
                    \frac{1}{12} & \frac{1}{12} & 0 \\
                    -\frac{1}{12} & -\frac{1}{12} & 0 \\
                  \end{array}
                \right)
  +(\epsilon^L_{13})^3\left(
                  \begin{array}{ccc}
                    0 & 0 & -\frac{1}{6}e^{-i\delta^L} \\
                    \frac{1}{12}e^{i\delta^L} & \frac{1}{12}e^{i\delta^L} & 0 \\
                    \frac{1}{12}e^{i\delta^L} & \frac{1}{12}e^{i\delta^L} & 0 \\
                  \end{array}
                \right)\nonumber\\
  &+&\epsilon^L_{12}(\epsilon^L_{13})^2\left(
                  \begin{array}{ccc}
                    \frac{\sqrt{2}}{4} & -\frac{\sqrt{2}}{4} & 0 \\
                    0 & 0 & 0 \\
                    0 & 0 & 0 \\
                  \end{array}
                \right)
  +(\epsilon^L_{12})^2\epsilon^L_{13}e^{i\delta^L}\left(
                  \begin{array}{ccc}
                    0 & 0 & 0 \\
                    \frac{1}{4} & \frac{1}{4} & 0 \\
                    \frac{1}{4} & \frac{1}{4} & 0 \\
                  \end{array}
                \right)
+\mathcal{O}\left((\epsilon^L_{ij})^4\right)\nonumber\\
&=&U_{\rm BM}
-\lambda\left(\begin{array}{ccc}  -\frac{\sqrt{2}}{2} & \frac{\sqrt{2}}{2} & -Ze^{-i2\delta^L} \\
                                           -\frac{1}{2}(1-Z) & -\frac{1}{2}(1-Z) & 0 \\
                                           \frac{1}{2}(1+Z) & \frac{1}{2}(1+Z) & 0 \\
                                           \end{array}
                                           \right)\nonumber\\
  &+&\lambda^2\left(
                  \begin{array}{ccc}
                    -\frac{\sqrt{2}}{4}(1+Z^2e^{-i2\delta^L}) &-\frac{\sqrt{2}}{4}(1+Z^2e^{-i2\delta^L})& 0 \\
                    -\frac{1}{2}(A+Z-\frac{1}{2}) & \frac{1}{2}(A+Z-\frac{1}{2}) & -\frac{\sqrt{2}}{2}(A+\frac{1}{2}Z^2e^{-i2\delta^L}) \\
                    -\frac{1}{2}(A+Z+\frac{1}{2}) & \frac{1}{2}(A+Z+\frac{1}{2}) & \frac{\sqrt{2}}{2}(A-\frac{1}{2}Z^2e^{-i2\delta^L}) \\
                  \end{array}
                \right)
  \nonumber\\
  &+&\lambda^3\left(
                  \begin{array}{ccc}
                    -\frac{\sqrt{2}}{4}(\frac{1}{3}+Z^2e^{-i2\delta^L}) & \frac{\sqrt{2}}{4}(\frac{1}{3}+Z^2e^{-i2\delta^L}) & -\frac{1}{6}e^{-i4\delta^L} \\
                    \frac{1}{12}(6AZ+6A+3Z+Z^3e^{-i2\delta^L}-1) & \frac{1}{12}(6AZ+6A+3Z+Z^3e^{-i2\delta^L}-1) & 0 \\
                    \frac{1}{12}(-6AZ+6A+3Z+Z^3e^{-i2\delta^L}+1) & \frac{1}{12}(-6AZ+6A+3Z+Z^3e^{-i2\delta^L}+1) & 0 \\
                  \end{array}
                \right)\nonumber\\
 & +&\mathcal{O}\left(\lambda^4\right).
\end{eqnarray}

\subsection{Expansion on tribimaximal pattern}

If we start expansion with an alternative TB mixing form, the
triminimal parametrization of the PMNS matrix is obtained
as~\cite{triminimalneutrino1,triminimalneutrino2} to the second order in $\epsilon_{ij}^L$:

\begin{eqnarray}
U_{\mathrm{PMNS}} &=&U_{\rm TB}
+ \epsilon^L_{12}\left(\begin{array}{ccc}  -\frac{1}{\sqrt{3}} & \frac{2}{\sqrt{6}}  & 0 \\
                                           -\frac{1}{\sqrt{3}} & -\frac{1}{\sqrt{6}} & 0 \\
                                           \frac{1}{\sqrt{3}}  & \frac{1}{\sqrt{6}}  & 0 \\
                                           \end{array}
                                           \right)
  +\epsilon^L_{23}\left(
                  \begin{array}{ccc}
                    0 & 0 & 0 \\
                    \frac{1}{\sqrt{6}} & -\frac{1}{\sqrt{3}} & \frac{1}{\sqrt{2}} \\
                    \frac{1}{\sqrt{6}} & -\frac{1}{\sqrt{3}} & -\frac{1}{\sqrt{2}} \\
                  \end{array}
                \right)
  +\epsilon^L_{13}\left(
                  \begin{array}{ccc}
                    0 & 0 & e^{-i\delta^L} \\
                    -\frac{1}{\sqrt{3}}e^{i\delta^L} & -\frac{1}{\sqrt{6}}e^{i\delta^L} & 0 \\
                    -\frac{1}{\sqrt{3}}e^{i\delta^L} & -\frac{1}{\sqrt{6}}e^{i\delta^L} & 0 \\
                  \end{array}
                \right)\nonumber\\
  &+&(\epsilon^L_{12})^2\left(
                  \begin{array}{ccc}
                    -\frac{1}{\sqrt{6}} & -\frac{1}{2\sqrt{3}} & 0 \\
                    \frac{1}{2\sqrt{6}} & -\frac{1}{2\sqrt{3}} & 0 \\
                    -\frac{1}{2\sqrt{6}} & \frac{1}{2\sqrt{3}} & 0 \\
                  \end{array}
                \right)
  +(\epsilon^L_{23})^2\left(
                  \begin{array}{ccc}
                    0 & 0 & 0 \\
                    \frac{1}{2\sqrt{6}} & -\frac{1}{2\sqrt{3}} & -\frac{1}{2\sqrt{2}} \\
                    -\frac{1}{2\sqrt{6}} & \frac{1}{2\sqrt{3}} & -\frac{1}{2\sqrt{2}} \\
                  \end{array}
                \right)
  +(\epsilon^L_{13})^2\left(
                  \begin{array}{ccc}
                    -\frac{1}{\sqrt{6}} & -\frac{1}{2\sqrt{3}} & 0 \\
                    0 & 0 & -\frac{1}{2\sqrt{2}} \\
                    0 & 0 & -\frac{1}{2\sqrt{2}} \\
                  \end{array}
                \right)\nonumber\\
  &+&\epsilon^L_{12}\epsilon^L_{23}\left(
                  \begin{array}{ccc}
                    0 & 0 & 0 \\
                    \frac{1}{\sqrt{3}} & \frac{1}{\sqrt{6}} & 0 \\
                    \frac{1}{\sqrt{3}} & \frac{1}{\sqrt{6}} & 0 \\
                  \end{array}
                \right)
  +\epsilon^L_{12}\epsilon^L_{13}e^{i\delta^L}\left(
                  \begin{array}{ccc}
                    0 & 0 & 0 \\
                    \frac{1}{\sqrt{6}} & -\frac{1}{\sqrt{3}} & 0 \\
                    \frac{1}{\sqrt{6}} & -\frac{1}{\sqrt{3}} & 0 \\
                  \end{array}
                \right)
  +\epsilon^L_{23}\epsilon^L_{13}e^{i\delta^L}\left(
                  \begin{array}{ccc}
                    0 & 0 & 0 \\
                    -\frac{1}{\sqrt{3}} & -\frac{1}{\sqrt{6}} & 0 \\
                    \frac{1}{\sqrt{3}} & \frac{1}{\sqrt{6}} & 0 \\
                  \end{array}
                \right)+{\cal O}((\epsilon_{ij}^L)^3),\label{prw}
\end{eqnarray}
where $\epsilon^L_{23}$, $\epsilon^L_{13}$, and $\delta^L$ are the
same parameters as parametrization of the PMNS matrix just like in the
BM triminimal expansion case, whereas $\epsilon^L_{12}$ is not. This
set of expansion parameters is certainly better than the one in the
previous section if convergency is the criteria of the expansion.
With the replacements $\epsilon^L_{12}= B \lambda^3$,
$\epsilon^L_{23}=-A \lambda^2$ and
$\epsilon^L_{13}e^{i\delta^L}=\lambda Z$, the hierarchy is
$\epsilon_{12}^L\sim (\epsilon_{13}^L)^3$ and $\epsilon_{23}^L\sim
(\epsilon_{13}^L)^2$. The parameters $A$ and $\lambda$ are the same
Wolfenstein parameters as those in the BM case, argued from the aspect of QLC
relations, and $B$ is a new parameter, which is of order ${\cal
O}(1)$, adjusted to fit the data according to the estimate in
Sec.~II. Thus to the third order of $\epsilon_{13}^L$:

\begin{eqnarray}
U_{\mathrm{PMNS}} &=&U_{\rm TB}
+ \epsilon^L_{12}\left(\begin{array}{ccc}  -\frac{1}{\sqrt{3}} & \frac{2}{\sqrt{6}}  & 0 \\
                                           -\frac{1}{\sqrt{3}} & -\frac{1}{\sqrt{6}} & 0 \\
                                           \frac{1}{\sqrt{3}}  & \frac{1}{\sqrt{6}}  & 0 \\
                                           \end{array}
                                           \right)
  +\epsilon^L_{23}\left(
                  \begin{array}{ccc}
                    0 & 0 & 0 \\
                    \frac{1}{\sqrt{6}} & -\frac{1}{\sqrt{3}} & \frac{1}{\sqrt{2}} \\
                    \frac{1}{\sqrt{6}} & -\frac{1}{\sqrt{3}} & -\frac{1}{\sqrt{2}} \\
                  \end{array}
                \right)
  +\epsilon^L_{13}\left(
                  \begin{array}{ccc}
                    0 & 0 & e^{-i\delta^L} \\
                    -\frac{1}{\sqrt{3}}e^{i\delta^L} & -\frac{1}{\sqrt{6}}e^{i\delta^L} & 0 \\
                    -\frac{1}{\sqrt{3}}e^{i\delta^L} & -\frac{1}{\sqrt{6}}e^{i\delta^L} & 0 \\
                  \end{array}
                \right)\nonumber\\
&+&(\epsilon^L_{13})^2\left(
                  \begin{array}{ccc}
                    -\frac{1}{\sqrt{6}} & -\frac{1}{2\sqrt{3}} & 0 \\
                    0 & 0 & -\frac{1}{2\sqrt{2}} \\
                    0 & 0 & -\frac{1}{2\sqrt{2}} \\
                  \end{array}
                \right)
  +\epsilon^L_{23}\epsilon^L_{13}e^{i\delta^L}\left(
                  \begin{array}{ccc}
                    0 & 0 & 0 \\
                    -\frac{1}{\sqrt{3}} & -\frac{1}{\sqrt{6}} & 0 \\
                    \frac{1}{\sqrt{3}} & \frac{1}{\sqrt{6}} & 0 \\
                  \end{array}
                \right)  +{\cal O}((\epsilon_{13}^L)^4)           \nonumber\\
 &=&U_{\rm TB}
  +\lambda\left(
                  \begin{array}{ccc}
                    0 & 0 & e^{-i2\delta^L} \\
                    -\frac{1}{\sqrt{3}}Z & -\frac{1}{\sqrt{6}}Z & 0 \\
                    -\frac{1}{\sqrt{3}}Z & -\frac{1}{\sqrt{6}}Z & 0 \\
                  \end{array}
                \right)     -\lambda^2\left(
                  \begin{array}{ccc}
                    \frac{1}{\sqrt{6}}Z^2e^{-i2\delta^L} & \frac{1}{2\sqrt{3}}Z^2e^{-i2\delta^L} & 0 \\
                    \frac{1}{\sqrt{6}}A& -\frac{1}{\sqrt{3}}A & \frac{1}{2\sqrt{2}}(2A+Z^2e^{-i2\delta^L}) \\
                    \frac{1}{\sqrt{6}}A & -\frac{1}{\sqrt{3}}A & -\frac{1}{2\sqrt{2}}(2A-Z^2e^{-i2\delta^L}) \\
                  \end{array}\right) \nonumber\\
   &+&\lambda^3\left(
                  \begin{array}{ccc}
                    -\frac{1}{\sqrt{3}}B & \frac{2}{\sqrt{6}}B & 0 \\
                    -\frac{1}{\sqrt{3}}(B-AZ)& -\frac{1}{\sqrt{6}}(B-AZ) & 0 \\
                    \frac{1}{\sqrt{3}}(B-AZ) & \frac{1}{\sqrt{6}}(B-AZ) & 0 \\
                  \end{array}
                \right)  +{\cal O}(\lambda^4).
                \end{eqnarray}

It is novel that the deviations of the lepton sector are explicitly
illustrated in orders of Wolfenstein parameter $\lambda$ from the
quark sector, along with the idea to understand both the quark and
lepton mixing patterns in a unified
manner~\cite{triminimalquark,triminimalquark2}.

\section{A phenomenological proposal of new mixing pattern}\label{section:new}
From previous global fits of neutrino mixing angles,
\begin{eqnarray}
\theta_{12}^{\nu}\simeq 33.957^{\circ +2.434^\circ}_{~-2.143^\circ},~~~\theta_{23}^{\nu}\simeq 43.050^{\circ +7.839^\circ}_{~-5.834^\circ},
\end{eqnarray}
together with the latest T2K implication:
\begin{eqnarray}
\theta_{13}^{\nu}=9.685^{\circ +4.698^\circ}_{~-6.289^\circ}~~{(\rm NH})~~~~{\rm or}~~\theta_{13}^{\nu}=10.986^{\circ+5.218^\circ}_{~-6.848^\circ}~~({\rm IH}),
\end{eqnarray}
we can find a self-complementary relation between the mixing angles
\begin{eqnarray}
\theta_{12}^{\nu} + \theta_{13}^{\nu} \simeq \theta_{23}^{\nu} \simeq 45^\circ.
\end{eqnarray}
 This leads to a proposal for a new mixing pattern which is closer to the experimental data than BM and TM patterns.
To construct such a new mixing pattern, we begin with the assumption that
\begin{eqnarray}
\nonumber
&{\rm sin}^2\theta_{23}=\frac{1}{2},&\\
\nonumber
&{\rm sin}^2\theta_{12}=\frac{1}{3},&\\
&{\rm sin}^2\theta_{13}={\rm
sin}^2(\theta_{23}-\theta_{12})=\frac{1}{2}-\frac{\sqrt{2}}{3}.&
\label{newangle}
\end{eqnarray}
Thus the new mixing matrix could be given as
\begin{eqnarray}
U_{\rm NM}&=&\left(
\begin{array}{ccc}
\frac{\sqrt{2}}{\sqrt{3}}\sqrt{\frac{1}{2}+\frac{\sqrt{2}}{3}}& \frac{1}{\sqrt{3}}\sqrt{\frac{1}{2}+\frac{\sqrt{2}}{3}} & \frac{1}{\sqrt{3}}-\frac{1}{\sqrt{6}} \\
-\frac{1}{3}+\frac{\sqrt{2}}{6}-\frac{1}{\sqrt{6}} & -\frac{\sqrt{2}}{6}+\frac{1}{6}+\frac{1}{\sqrt{3}} & \frac{\sqrt{2}}{2}\sqrt{\frac{1}{2}+\frac{\sqrt{2}}{3}} \\
-\frac{1}{3}+\frac{\sqrt{2}}{6}+\frac{1}{\sqrt{6}} & -\frac{\sqrt{2}}{6}+\frac{1}{6}-\frac{1}{\sqrt{3}} & \frac{\sqrt{2}}{2}\sqrt{\frac{1}{2}+\frac{\sqrt{2}}{3}}
\end{array}\right)\nonumber\\
&\simeq&\left(
\begin{array}{ccc}
0.8047& 0.5690& 0.1691 \\
-0.5059 &0.5083  & 0.6969 \\
 0.3106& -0.6464 & 0.6969
\end{array}\right).
\end{eqnarray}

One can easily see that for this new mixing pattern, the moduli of
the mixing matrix elements are compatible with the results from
global fits~\cite{zheng,nu_data1,nu_data2}. Generally, the mixing
matrix is expanded as below. With the replacements
$\epsilon_{12}^L=B\lambda^3$, $\epsilon_{23}^L=-A\lambda^2$, and
$\epsilon_{13}^Le^{i\delta_L}=\lambda^3Z^\prime$ or
$\lambda^2Z^\prime$ defined according to the hierarchical structure,
the expansion could be obtained to the order of ${\cal
O}(\lambda^3)$:
\begin{eqnarray}
U_{\rm PMNS}&=&U_{\rm NM} +\epsilon_{12}\left( \begin{array}{ccc}
-s_{12}^0c_{13}^0 & c_{12}^0c_{13}^0 &0
\\ s_{12}^0s_{13}^0s_{23}^0e^{i\delta}-c_{12}^0c_{23}^0 & -c_{12}^0s_{13}^0s_{23}^0e^{i\delta}-s_{12}^0c_{23}^0 & 0 \\
s_{12}^0s_{13}^0c_{23}^0e^{i\delta}+c_{12}^0s_{23}^0 & -c_{12}^0s_{13}^0c_{23}^0e^{i\delta}+s_{12}^0s_{23}^0 &0
\end{array}\right)\nonumber\\
&+&\epsilon_{23}\left(\begin{array}{ccc}
0&0&0\\
-s_{13}^0c_{12}^0c_{23}^0e^{i\delta}+s_{23}^0s_{12}^0 &-s_{12}^0s_{13}^0c_{23}^0e^{i\delta}-s_{23}^0c_{12}^0 &c_{13}^0c_{23}^0\\
c_{12}^0s_{23}^0s_{13}^0e^{i\delta}+c_{23}^0s_{12}^0
&s_{12}^0s_{23}^0s_{13}^0e^{i\delta}-c_{12}^0c_{23}^0&c_{12}^0s_{23}^0
\end{array}\right)\nonumber\\
&+&\epsilon_{13}\left( \begin{array}{ccc}
-c_{12}^0s_{13}^0&-s_{12}^0s_{13}^0&c_{13}^0e^{-i\delta}\\
-c_{12}^0c_{13}^0s_{23}^0e^{i\delta}&-c_{13}^0s_{23}^0s_{12}^0e^{i\delta}&-s_{13}^0s_{23}^0\\
-c_{12}^0c_{13}^0c_{23}^0e^{i\delta}&-s_{12}^0c_{13}^0c_{23}^0e^{i\delta}&-s_{13}^0c_{23}^0
\end{array}\right)+\mathcal {O}(\lambda^4)\;, \label{general}
\end{eqnarray}
where $s_{ij}^0=\sin\theta_{ij}$, $c_{ij}^0=\cos\theta_{ij}$ with
$\theta_{ij}$ denotes the complementary mixing angles proposed in
Eq.~(\ref{newangle}). We see that this expansion based on the new
mixing pattern is much simpler compared with the triminimal
expansions based on either BM or TM mixing patterns in the above
section. The leading order basis matrix $U_{\rm NM}$ is much closer
to the experimental fits.

\section{Summary}\label{section:summary}

In summary, we re-analyzed the neutrino mixing patterns according to
the latest T2K result for a larger mixing angle $\theta_{13}$, and
found that the deviation of the neutrino mixing pattern from
bimaximal (BM) and tribimaximal (TB) patterns become larger than
previously expected. We also examined relations connecting quark and
lepton mixing matrices and it turned out that several schemes can
still accommodate the latest T2K result nicely. As a necessary
update of previous works, we made new triminimal expansions of the
PMNS mixing matrix based on BM and TB mixing patterns in terms of a
more reasonable $\lambda$ hierarchy. From the phenomenological point
of view, we also proposed a new mixing pattern with a
self-complementary relation between the mixing angles
$\theta_{12}^{\nu} + \theta_{13}^{\nu} \simeq 45^\circ$, and
concluded that such a mixing pattern works well in providing rather
good descriptions to the data at least with a precision of $\mathcal
{O}(\lambda^2)$. Thus the new T2K result can enrich our
understanding of neutrino properties.

~

~

This work is partially supported by National Natural Science
Foundation of China (Grants No.~11021092, No.~10975003,
No.~11035003, and No.~11120101004) and by Peking University Visiting
Scholar Program for Graduate Students.

%
%
%

%
%

\end{onecolumn}

\end{document}